\documentclass[graybox,natbib,nosecnum]{svmult}
\bibpunct{(}{)}{;}{a}{}{,} 

\pdfoutput=1   

\usepackage{mathptmx}       
\usepackage{helvet}         
\usepackage{courier}        
\usepackage{type1cm}        

\usepackage{makeidx}         
\usepackage{graphicx}        
\usepackage{multicol}        
\usepackage[bottom]{footmisc}
\usepackage[normalem]{ulem}	
\usepackage{hyperref}  

\usepackage{soul}   

\makeindex             

\begin{document}

\title*{Brown Dwarf Formation: Theory}
\author{Anthony P. Whitworth}
\institute{Anthony P. Whitworth \at School of Physics \& Astronomy, Cardiff University, The Parade, Cardiff CF24 3AA, Wales, UK, \email{A.Whitworth@astro.cf.ac.uk}}
%
%
\maketitle

\abstract{We rehearse the physical and theoretical considerations that define the nature of Brown Dwarfs, in particular the maximum mass for a Brown Dwarf (set by the Hydrogen-Burning Limit) and the minimum mass for a Brown Dwarf (set by the Opacity Limit). We then review the range of mechanisms that have been invoked to explain the formation of Brown Dwarfs and their statistical properties. These include turbulent fragmentation, fragmentation of filaments and discs, dynamical ejection of stellar embryos, and photoerosion. The primary contenders would seem to be turbulent fragmentation and disc fragmentation, and the observations needed to evaluate their relative importance may soon be available.}

\section{Introduction }

To avoid ambiguity, we define Brown Dwarfs (BDs) as self-gravitating bodies which (a) have sufficiently low mass that they can be held up by the electron-degeneracy pressure of a gas that is too cool to support hydrogen burning, and (b) have condensed out of interstellar gas on a dynamical timescale ($\stackrel{<}{\sim}10^5\,{\rm years}$), due to gravitational instability, and therefore with an essentially uniform elemental composition. Condition (a) means that Brown Dwarfs have masses below the Hydrogen-Burning Limit at $M_{_{\rm HB.LIM}}\sim 0.075\pm 0.005\,{\rm M}_{_\odot}$; the precise value of $M_{_{\rm HB.LIM}}$ depends weakly on the elemental composition of the gas forming the Brown Dwarf. Condition (b) means that Brown Dwarfs are more closely related to stars than to planets (that is, if one presumes that planets form by core accretion, on a much longer timescale, $\stackrel{>}{\sim}10^6\,{\rm years}$), {\it and} that Brown Dwarfs cannot have masses below the Opacity Limit at $M_{_{\rm OP.LIM}}\sim 0.003\pm 0.001\,{\rm M}_{_\odot}$. This distinction between stars and planets is a useful theoretical perspective, based on three perceptions. First, the Initial Mass Function for star formation, and the associated binary statistics, whilst poorly constrained at low masses, appear to be continuous across the Hydrogen-Burning Limit \citep{Kroupa2002,Chabrier2003}. At the very least, this suggests that whatever mechanisms are involved in the formation of low-mass hydrogen-burning stars, these same mechanisms are also involved in the formation of high-mass Brown Dwarfs. Second, the processes determining the mass of a star are presumed to occur at densities $\stackrel{<}{\sim}10^{-13}\,{\rm g}\,{\rm cm}^{-3}$ and temperatures $\stackrel{<}{\sim}100\,{\rm K}$ \citep[or possibly $\stackrel{<}{\sim}10^{-4}\,{\rm g}\,{\rm cm}^{-3}$ and $\stackrel{<}{\sim}5000\,{\rm K}$, if there is a second fragmentation stage due to the increase in the specific heat of the gas when molecular hydrogen is being dissociated, but see][]{Bate1998}, i.e. long before the condensing matter reaches the densities and temperatures at which electron degeneracy or hydrogen burning have any influence. Third, current theories of star formation do not preclude the formation of stars with masses below the Hydrogen-Burning Limit. The minimum mass for star formation is normally identified with the Opacity Limit at $M_{_{\rm OP.LIM}}\sim 0.003\!\pm\! 0.001\,{\rm M}_{_\odot}$, which is the smallest gravitationally bound mass that is able to cool radiatively on a dynamical timescale; it is therefore the smallest fragment that cannot break up gravitationally into even smaller fragments, and it is evidently much smaller than the Hydrogen-Burning Limit. Note that, by extension, our definition of Brown Dwarfs includes objects below the Deuterium-Burning Limit at $M_{_{\rm DB.LIM}}\sim 0.0125\pm0.005\,{\rm M}_{_\odot}$, provided they form by gravitational instability.

In the next two sections, we introduce some fundamental definitions and equations, plot the key stages in the evolution of a prestellar core (i.e. a lump of interstellar gas that is gravitationally unstable and therefore destined to for a star, or tight cluster of stars), and derive the Opacity Limit for dynamical star formation (which gives the minimum mass for a Brown Dwarf) and the Hydrogen-Burning Limit (which gives the maximum mass for a Brown Dwarf). In the following five sections, we consider the formation of prestellar cores of Brown-Dwarf mass (i)  by turbulent fragmentation; (ii) by fragmentation of the filaments feeding material into  massive clusters; (iii) by fragmentation of protostellar discs; (iv) when small protostellar embryos are ejected dynamically from their birth clusters; and (v) from massive cores that are photo-eroded when they become embedded in an H{\sc ii} region. The final two sections discuss the limitations of the results we have presented, and summarise our main conclusions.

\section{Preliminaries}

Under the circumstances with which we shall be concerned, the global evolution of an approximately spherical , {\em uniform-density} core depends on the density of self-gravitational potential energy, $P_{_{\rm GRAV}}$. For a uniform-density prestellar core,
\begin{eqnarray}\label{EQN:P_grav}
P_{_{\rm GRAV}}&=&-\;\frac{(4\pi)^{1/3}\,G\,(3M)^{2/3}}{5}\;\rho^{4/3}\,;
\end{eqnarray}
we give it the symbol $P_{_{\rm GRAV}}$, because it has the dimensions of pressure, but it acts like a negative pressure, pulling inwards.

Self-gravity is resisted by internal pressure, which at interstellar densities means ideal gas pressure, 
\begin{eqnarray}\label{EQN:P_ig}
P_{_{\rm IG}}&=&\frac{\rho\, k_{_{\rm B}}\, T}{\bar m}\;\,=\;\,\rho\, a^2\,.
\end{eqnarray}
However, as the prestellar core approaches stellar densities, there may also be a significant contribution from electron-degeneracy pressure, which, in the non-relativistic limit relevant to very low-mass stars, takes the form
\begin{eqnarray}\label{EQN:P_nred}
P_{_{\rm NRED}}&=&\frac{\pi\, h^2}{60\,m_{_{\rm e}}}\left(\!\frac{3}{\pi\,{\bar m}_{_{\rm e}}}\!\right)^{\!5/3}\,\rho^{5/3}\,.
\end{eqnarray}
Most variables here have their standard identities, but we note that ${\bar m}$ is the mean gas-particle mass (which, for gas with solar composition, has the values ${\bar m}\simeq 4\times 10^{-24}\,{\rm g}$ when the hydrogen is molecular, and ${\bar m}\simeq 10^{-24}\,{\rm g}$ when the gas is fully ionised); 
\begin{eqnarray}\label{EQN:a_iso}
a&=&\left(\frac{k_{_{\rm B}}\,T}{\bar m}\right)^{1/2}
\end{eqnarray}
is the isothermal sound speed; $m_{_{\rm e}}$ is the mass {\em of} an electron; and ${\bar m}_{_{\rm e}}$ is the mass {\em associated with} an electron when account is taken of the other particle species present (for fully ionised gas with solar composition, ${\bar m}_{_{\rm e}}\simeq 2\times 10^{-24}\,{\rm g}$).

The core's evolution depends on the sign of
\begin{eqnarray}\label{EQN:P_eff}
P_{_{\rm EFF}}=&P_{_{\rm GRAV}}+3P_{_{\rm IG}}+3P_{_{\rm NRED}}\,.
\end{eqnarray}
If $P_{_{\rm EFF}}<0$, the core contracts; if $P_{_{\rm EFF}}=0$, the core is in hydrostatic equilibrium; and if $P_{_{\rm EFF}}>0$, the core expands. The factor 3 derives from the Virial Theorem, and reflects the fact that both $P_{_{\rm IG}}$ and $P_{_{\rm NRED}}$ are delivered by non-relativistic particles. Self-gravity may also be resisted by rotational and/or magnetic stresses, but, when these are important, the core departs significantly from spherical symmetry, and so these cases must be treated separately (for example, as in the treatment of disc fragmentation below).

If $|P_{_{\rm GRAV}}|\gg3P_{_{\rm IG}}+3P_{_{\rm NRED}}$ (i.e. negligible internal pressure), a spherical uniform-density core collapses to a point on a freefall timescale, 
\begin{eqnarray}
t_{_{\rm FF}}&=&\left(\!\frac{3\,\pi}{32\,G\,\rho}\!\right)^{\!1/2}\,.
\end{eqnarray}

It will frequently be convenient to work with the isothermal sound speed, $a$, rather than the temperature, $T$. Thus, for example, using Eqn. (\ref{EQN:a_iso}) the flux from a blackbody like surface can be written
\begin{eqnarray}
F_{_{\rm BB}}&=&\sigma_{_{\rm SB}}\,T^4\;\,\equiv\;\,\frac{2\,\pi^5\,k_{_{\rm B}}^4\,T^4}{15\,c^2\,h^3}\;\,=\;\,\frac{2\,\pi^5\,{\bar m}^4\,a^8}{15\,c^2\,h^3}\,.
\end{eqnarray}
In the same vein, the Rosseland- and Planck-mean opacities for interstellar gas with solar composition -- in the temperature range, $3\,{\rm K}\stackrel{<}{\sim} T \stackrel{<}{\sim} 300\,{\rm K}$, where they are dominated by dust -- can be approximated by 
\begin{eqnarray}\label{EQN:kappa}
{\bar\kappa}&\simeq&\kappa_{_T}\,T^2\;\,\simeq\;\,\kappa_{_a}\,a^4\,.
\end{eqnarray}
with $\;\kappa_{_T}\simeq7\times10^{-4}\,{\rm cm}^2\,{\rm g}^{-1}\,{\rm K}^{-2}\;$ and $\;\kappa_{_a}\simeq6\times 10^{-19}\,{\rm s}^4\,{\rm cm}^{-2}\,{\rm g}^{-1}$ \citep[see][]{Whitwort2016}.

\section{The condensation of a prestellar core}

In this section we consider the processes that occur between when a prestellar core is formed and starts to contract, and when the core reaches stellar densities. This enables us to estimate the minimum and maximum mass for a Brown Dwarf, and the rest of the review is then concerned with the processes that form prestellar cores of Brown Dwarf mass.

A prestellar core of mass $M$ can only contract to become a star if its self-gravity overcomes its internal pressure, i.e. $P_{_{\rm EFF}}<0$ (see Eqn. \ref{EQN:P_eff}). At the low densities obtaining in interstellar space, we can ignore electron-degeneracy pressure, so this condition becomes $P_{_{\rm GRAV}}+3P_{_{\rm IG}}<0$, or 
\begin{eqnarray}\label{EQN:rho_1}
\rho&\stackrel{>}{\sim}&\rho_{_{\rm JEANS}}\;\,\simeq\;\,\frac{30\,a^6}{G^3\,M^2}\;\,\longrightarrow\;\,1.6\times 10^{-18}\,{\rm g}\,{\rm cm}^{-3}\left(\!\frac{M}{{\rm M}_{_\odot}}\!\right)^{\!-2},\\\label{EQN:rho_2}
n_{_{{\rm H}_2}}\!&\stackrel{>}{\sim}&n_{_{{\rm H_2.JEANS}}}\;\,\simeq\;\,\frac{30\,a^6}{G^3\,M^2\,{\bar m}_{_{{\rm H}_2}}}\;\,\longrightarrow\;\,\;\,3.2\times 10^5\,{\rm H}_2\,{\rm cm}^{-3}\left(\!\frac{M}{{\rm M}_{_\odot}}\!\right)^{-2},
\end{eqnarray}
The last expression on the righthand side of Eqn. (\ref{EQN:rho_1}) is obtained by substituting typical values for local low-mass star formation regions (hereafter `local values'), viz. $T\simeq10\,{\rm K}$, ${\bar m}\simeq4\times 10^{-24}\,{\rm g}$ (corresponding to gas with solar elemental composition in which the hydrogen is molecular), and hence $a\simeq 0.2\,{\rm km}\,{\rm s}^{-1}$. Eqn. (\ref{EQN:rho_2}) gives the corresponding number-density of hydrogen molecules, and is derived from Eqn. (\ref{EQN:rho_1}) using the fact that for molecular gas with solar composition, the mass associated with one hydrogen molecule is ${\bar m}_{_{{\rm H}_2}}=5\times 10^{-24}\,{\rm g}$, when account is taken of other elemental species, in particular helium. The equivalent constraint on the radius is
\begin{eqnarray}\label{EQN:R_Jeans}
R\;\,\stackrel{<}{\sim}\;\,R_{_{\rm JEANS}}&\simeq&\frac{G\,M}{5\,a^2}\;\,\longrightarrow\;\,4500\,{\rm AU}\,\left(\frac{M}{{\rm M}_{_\odot}}\right)\,.
\end{eqnarray}
Evidently prestellar cores of Brown Dwarf mass are very dense ($\stackrel{>}{\sim}5\times 10^7\,{\rm H}_{_2}\,{\rm cm}^{-3}$) and small ($\stackrel{<}{\sim}\,400\,{\rm AU}$) at their inception.

A core of mass $M$ satisfying Eqns (\ref{EQN:rho_1}) through (\ref{EQN:R_Jeans}) will start to contract. Moreover, as long as the contracting core is able to radiate away the work done by compression, the gas remains approximately isothermal ($a$ approximately constant) and $|P_{_{\rm GRAV}}|$ increases faster than $P_{_{\rm IG}}$ (with increasing $\rho$), so the contraction accelerates and approaches freefall collapse. Under this circumstance, departures from spherical symmetry (lumpiness, flattening) tend to be amplified, because Eqn. (\ref{EQN:rho_1}) becomes satisfied by smaller and smaller masses. Consequently the core can fragment into smaller cores, and thereby spawn more than one star.

The radiative cooling rate for a core is given by
\begin{eqnarray}\label{EQN:C_rad}
{\cal C}_{_{\rm RAD}}&\simeq&\frac{4\,\pi\, R^2\,\sigma_{_{\rm SB}}\,T^4}{\left({\bar \tau}_{_{\rm ROSSELAND}}\,+\,{\bar \tau}_{_{\rm PLANCK}}^{-1}\right)}\;\,\longrightarrow\;\,\frac{4\,\pi^6\,{\bar m}^4\,a^8}{15\,c^2\,h^3}\left(\!\frac{3\,M}{4\,\pi\,\rho}\!\right)^{\!2/3}\!.\hspace{0.7cm}
\end{eqnarray}
Here ${\bar \tau}_{_{\rm ROSSELAND}}$ is the Rosseland-mean opacity between the centre of the core and its surface, ${\bar \tau}_{_{\rm PLANCK}}$ is the corresponding Planck-mean opacity. The final expression on the righthand side of Eqn. (\ref{EQN:C_rad}) is obtained by setting $({\bar \tau}_{_{\rm ROSSELAND}}+{\bar \tau}_{_{\rm PLANCK}}^{-1})\sim 2$, which is equivalent to assuming that the core is marginally optically thick and therefore radiates approximately like a blackbody. It is straightforward to relax this assumption, but the algebra becomes more cumbersome, and the result is not significantly changed \citep[as shown by][]{WhitStam2006}.

The compressional heating rate for a core in freefall is
\begin{eqnarray}\label{EQN:H_comp}
{\cal H}_{_{\rm COMP.FF}}&=&\left(P\,\frac{dV}{dt}\right)_{_{\rm FF}}\;\,\simeq\;\,\frac{P\,V}{t_{_{\rm FF}}}\;\,\simeq\;\,\frac{M\,a^2}{t_{_{\rm FF}}}\;\,\simeq\;\,M\,a^2\left(\!\frac{32\,G\,\rho}{3\,\pi}\!\right)^{\!1/2}\!,\hspace{0.6cm}
\end{eqnarray}
so the approximately isothermal phase -- and hence the possibility of fragmentation -- lasts as long as ${\cal C}_{_{\rm RAD}}>{\cal H}_{_{\rm COMP.FF}}$, which reduces to the condition
\begin{eqnarray}\label{EQN:rhoHI}
\rho\!&\!<\!&\!\rho_{_{\rm HEAT.UP}}\,\simeq\,\left(\frac{3\pi^{35}{\bar m}^{24}a^{36}}{2^{11}5^6c^{12}h^{18}G^3M^2}\right)^{\!1/7}\,\longrightarrow\,1.6\times 10^{-14}\,{\rm g}\,{\rm cm}^{-3}\,\left(\frac{M}{{\rm M}_{_\odot}}\right)^{\!-2/7}\!.\hspace{0.7cm}
\end{eqnarray}

If we compare the lower limit on the density for a core of mass $M$ to start condensing under its self-gravity (Eqn. \ref{EQN:rho_1}) with 
this upper limit on the density for fragmentation of a core of mass $M$ (Eqn. \ref{EQN:rhoHI}), we see that massive cores experience a long isothermal collapse phase, and therefore are likely to fragment. For example, the density in a core of mass $1\,{\rm M}_{_\odot}$ increases by a factor of order $10^4$ during the isothermal phase, and its radius decreases by a factor of order 20; the density in a core of mass $10\,{\rm M}_{_\odot}$ increase by a factor of $5\times 10^5$ during the isothermal phase, and its radius decreases by a factor of order 80. In contrast,  lower-mass prestellar cores experience a very short isothermal collapse phase, and the minimum mass for star formation is defined as the mass of a core that has no isothermal phase between $\rho_{_{\rm JEANS}}$ and $\rho_{_{\rm HEAT.UP}}$, and therefore cannot fragment at all. Setting $\rho_{_{\rm JEANS}}=\rho_{_{\rm HEAT.UP}}$ and substituting from Eqns. (\ref{EQN:rho_1}) and (\ref{EQN:rhoHI}), we obtain the minimum mass for star formation, due to the Opacity Limit,
\begin{eqnarray}
M&>&M_{_{\rm OP.LIM}}\;\simeq\;\frac{5\,(5\pi)^{1/12}\,(24)^{1/2}\,m_{_{\rm PLANCK}}^3}{\pi^3\,{\bar m}^2}\left(\frac{k_{_{\rm B}}T}{{\bar m}c^2}\right)^{1/4}\,.
\end{eqnarray}
where $m_{_{\rm PLANCK}}=(hc/G)^{1/2}\simeq 5.5\times 10^{-5}\,{\rm g}$ is the Planck Mass \citep{Rees1976}

We note that the minimum mass due to the Opacity Limit is only weakly dependent on the temperature ($\propto T^{1/4}$), but quite strongly dependent on the mean gas-particle mass ($\propto{\bar m}^{-9/4}$). For contemporary star formation in the solar vicinity, where $T\simeq 10\,{\rm K}$ and ${\bar m}\simeq 4\times 10^{-24}\,{\rm g}$ (local conditions), 
\begin{eqnarray}
M_{_{\rm OP.LIM.LOCAL}}&\simeq &0.003\,{\rm M}_{_\odot}\,.
\end{eqnarray}
This is only of order three Jupiter masses, and suggests that there may well be an overlap in the mass ranges occupied by stars and planets. Distinguishing the nature of such objects will therefore require more information than just their masses. In contrast, for primordial star formation, the temperature was probably quite a lot higher (certainly higher than the then cosmic microwave background), say $T\simeq 500\,{\rm K}$, and the mean gas-particle mass was probably quite a lot lower, say ${\bar m}\simeq 2.5\times 10^{-24}\,{\rm g}$, giving
\begin{eqnarray}
M_{_{\rm OP.LIM.PRIMORDIAL}}&\simeq &0.023\,{\rm M}_{_\odot}\,;
\end{eqnarray}
this means that there was a smaller mass range for Brown Dwarfs forming in the early Universe.

For a core with mass $M>M_{_{\rm OP.LIMIT}}$, there is an isothermal phase, which starts when the core first becomes gravitationally unstable at $\rho_{_{\rm JEANS}}$, and ends at $\rho_{_{\rm HEAT.UP}}$ when it can no longer radiate fast enough. After this, there is a brief phase lasting 
\begin{eqnarray}
t_{_{\rm HEAT.UP}}&\sim&\left(\!\frac{3\pi}{32G\rho_{_{\rm HEAT.UP}}}\!\right)^{\!1/2}\;\,\longrightarrow\;\,150\,{\rm yr}\left(\!\frac{M}{{\rm M}_{_\odot}}\!\right)^{1/7}\,,
\end{eqnarray}
during which the density increases by about an order of magnitude, and the core heats up until it is close to hydrostatic balance ($\rho\simeq\rho_{_{\rm JEANS}}$; Eqn. \ref{EQN:rho_1}) and therefore approximately spherically symmetric. From thereon, contraction of the core continues, but it is mainly slow and quasistatic. As the core contracts, it becomes denser and hotter, in such a way as to remain close to hydrostatic balance, and this is called Kelvin-Helmholtz Contraction. The only exception to this quasistatic contraction occurs when the density reaches $\sim 10^{-7}\,{\rm g}\,{\rm cm}^{-3}$ and the temperature reaches $\sim 2000\,{\rm K}$, the specific heat of the gas rises, due to the dissociation of molecular hydrogen, and the core undergoes a second brief collapse phase. The slow contraction then resumes, and continues until either the gas becomes hot enough to sustain significant hydrogen burning, in which case it is a Main Sequence star, or it becomes dense enough to be supported by electron degeneracy pressure, in which case it is a Brown Dwarf. 

For low-mass cores (i.e. cores that will become low-mass stars), the rate of Kelvin-Helmholtz Contraction is determined by the rate at which half the self-gravitational potential energy being released can diffuse radiatively to the surface and escape. Hydrostatic balance is dominated by self-gravity, ideal-gas pressure, and non-relativistic electron-degeneracy pressure, i.e.
\begin{eqnarray}\label{EQN:KHC_1}
P_{_{\rm GRAV}}+3P_{_{\rm IG}}+3P_{_{\rm NRED}}&\simeq&0\,.
\end{eqnarray}
Substituting into Eqn. (\ref{EQN:KHC_1}) for $P_{_{\rm GRAV}}$ from Eqn. (\ref{EQN:P_grav}), for $P_{_{\rm IG}}$ from Eqn. (\ref{EQN:P_ig}), and for $P_{_{\rm NRED}}$ from Eqn. (\ref{EQN:P_nred}), we obtain the track of the core on the $(\rho,T)$ plane,
\begin{eqnarray}\label{EQN:KHC_2}
\frac{k_{_{\rm B}}\,{\bar T}}{\bar m}&=&\left(\frac{4\,\pi}{3}\right)^{\!1/3}\,\frac{G\,M^{2/3}}{5}\,{\bar \rho}^{\,1/3}\;\;-\;\;\frac{\pi\,h^2}{60\,m_{_{\rm e}}}\,\left(\frac{3}{\pi\,{\bar m}_{_{\rm e}}}\right)^{\!5/3}{\bar \rho}^{\,2/3}\,.
\end{eqnarray}
Because, significant temperature and density gradients are required to drive the diffusion (radiative, conductive or convective) that transports the core's luminosity to its surface, and also to hold the core up, ${\bar T}$ and ${\bar \rho}$ should be interpreted as mean values; the central values will be higher.

In the early stages of Kelvin-Helmholtz Contraction, when the density is low,
\begin{eqnarray}
{\bar \rho}&\ll&{\bar \rho}_{_{\rm FINAL}}\;\,\simeq\;\,\frac{2^8\,(\pi\,G\,m_{_{\rm e}})^3\,{\bar m}_{_{\rm e}}^5\,M^2}{(3\,h^2)^3}\;\,\simeq\;\,9.1\times10^4\,{\rm g}\,{\rm cm}^{-3}\left(\!\frac{M}{{\rm M}_{_\odot}}\!\right)^2\,,
\end{eqnarray}
the electron-degeneracy term on the far righthand side of Eqn. (\ref{EQN:KHC_2}) is negligible compared with the terms representing ideal-gas pressure and self-gravity, and so the temperature increases monotonically as ${\bar T}\propto {\bar\rho}^{\,1/3}$.

However, as the density approaches ${\bar\rho}_{_{\rm FINAL}}$, electron degeneracy pressure becomes increasingly important, and at 
\begin{eqnarray}
{\bar\rho}&=&\frac{{\bar\rho}_{_{\rm FINAL}}}{8}\;\,\simeq\;\,1.1\times 10^4\,{\rm g}\,{\rm cm}^{-3}\left(\!\frac{M}{{\rm M}_{_\odot}}\!\right)^2\,,
\end{eqnarray}
the temperature reaches a maximum, 
\begin{eqnarray}\label{EQN:T_max}
{\bar T}_{_{\rm MAX}}&\sim&\left(\frac{2\,\pi}{3}\right)^{\!4/3}\,\frac{G^2\,m_{_{\rm e}}\,{\bar m}\,{\bar m}_{_{\rm e}}^{5/3}\,M^{4/3}}{5\,h^2\,k_{_{\rm B}}}
\;\,\sim\;\,2.8\times 10^7\,{\rm K}\,\left(\!\frac{M}{{\rm M}_{_\odot}}\!\right)^{\!4/3}\,.
\end{eqnarray}

The critical issue is then whether, before reaching this maximum temperature, the temperature becomes sufficiently high to support hydrogen burning. In this case, the core has reached the Main Sequence, so it stops contracting and relaxes to a stable configuration, in which its luminosity is supplied by hydrogen burning. Moreover, for the low stellar masses with which we are concerned here ($M\stackrel{<}{\sim}0.3\,{\rm M}_{_\odot}$), the hydrogen-burning luminosities are so low that hydrogen is consumed very slowly and their Main Sequence lifetimes are much longer than the current age of the Universe; this is in effect the end of the road for all existing stars in this mass range, even if they formed at very high redshift.

However, if the maximum temperature is insufficient to support hydrogen burning, contraction only ceases at $\rho =\rho_{_{\rm FINAL}}$. The core has become a Brown Dwarf, supported by electron degeneracy pressure. It slowly settles towards $\rho_{_{\rm FINAL}}$, and its temperature and luminosity decline in perpetuity. 

For significant hydrogen burning to occur in a low-mass Main Sequence star requires that the central temperature, $T_{_{\rm CEN}}$, exceed $T_{_{\rm HB.LIM}}\simeq 2\times 10^6\,{\rm K}$. We shall assume that the star can be modelled as a polytrope with index $n=3/2$ (which is appropriate if the star is convective), in which case $T_{_{\rm CEN}}=1.9\,{\bar T}$. As first shown by \citet{Kumar1963}, the requirement that  $T_{_{\rm CEN}}$ exceed ${\bar T}_{_{\rm MAX}}$ (Eqn. \ref{EQN:T_max}) then gives us the minimum mass for a hydrogen-burning star, and hence the maximum mass for a Brown Dwarf,
\begin{eqnarray}
M_{_{\rm HB.LIM}}&\simeq&\left(\frac{2.0\times 10^6\,{\rm K}\;(1.9)^{-1}}{2.8\times 10^7\,{\rm K}}\right)^{\!3/4}\,{\rm M}_{_\odot}\;\,\simeq\;\,0.085\,{\rm M}_{_\odot}\,.
\end{eqnarray}
Given that we have used a one-zone model, the accuracy of this result is somewhat fortuitous. However, the basic underlying physics in the one-zone model is correct, and it allows us to capture the fundamental dependencies that would be invisible in a computational derivation of this result. In a full computer simulation, the contraction of a core is non-homologous, which means that the centre reaches stellar densities while the outer layers are still contracting.

\section{Forming brown dwarfs by turbulent fragmentation}\label{SEC:TurbFrag}

In this section, we consider the process of turbulent fragmentation \citep{PadoNord2002,PadoNord2004}. In turbulent fragmentation, stars form wherever two or more turbulent flows collide and create a prestellar core, i.e. a lump of gas that is sufficiently massive, dense and quiescent that it condenses out as a star, or a small cluster of stars. The issue here is whether this process operates effectively on the very low mass scales required to deliver a significant population of isolated Brown Dwarfs. If one re-formulates Eqn. (\ref{EQN:rho_1}) in terms of pressure, it becomes 
\begin{eqnarray}
P&=&\rho\,a^2\;\,\stackrel{>}{\sim}\;\,\frac{30\,a^8}{G^3\,M^2}\,.
\end{eqnarray}
In the turbulent fragmentation scenario, this pressure is the ram pressure of the turbulent gas flows that converge to form the core,
\begin{eqnarray}
P_{_{\rm RAM}}&\simeq&\rho_{_{\rm INFLOW}}\,v_{_{\rm INFLOW}}^2\;\,\simeq\;\,n_{_{{\rm H}_2.{\rm INFLOW}}}\,{\bar m}_{_{{\rm H}_2}}\,v_{_{\rm INFLOW}}^2\,,
\end{eqnarray}
and therefore the formation of a gravitationally unstable prestellar core of mass $M$ requires inflowing gas with
\begin{eqnarray}
n_{_{{\rm H}_2.{\rm INFLOW}}}v_{_{\rm INFLOW}}^2\!&\!\stackrel{>}{\sim}\!&\!\frac{30a^8}{G^3M^2{\bar m}_{_{{\rm H}_2}}}\,\longrightarrow\,1.4\times 10^7\,{\rm H}_{_2}\,{\rm cm}^{-3}\left({\rm km}\,{\rm s}^{-1}\right)^2\left(\frac{M}{0.03{\rm M}_{_\odot}}\right)^{\!-2}\!.\hspace{0.7cm}
\end{eqnarray}
In other words, the formation of Brown Dwarfs by turbulent fragmentation requires very large ram pressures, which means colliding flows with very large densities and velocities. There are two further complications. 

First, unless the colliding gas flows converge on a focal point from all directions -- which is rather unlikely -- the compressed gas tends not to condense to stellar densities, but rather disperses. This has been demonstrated with numerical simulations by \citet{Lomaetal2016}. Two antiparallel colliding flows with sufficient mass to produce a prestellar core of Brown-Dwarf mass produce a shock-compressed layer, but the layer is not extensive enough to be gravitationally unstable, and the gas escapes sideways (parallel to the shock) and disperses; once the inflow terminates, the gas can also disperse in the direction perpendicular to the shock \citep[c.f.][]{StoneM1970a,StoneM1970b}. \citet{Lomaetal2016} show that only convergent flows in which all three Cartesian contributions to the velocity divergence are negative produce gravitationally unstable condensations that spawn Brown Dwarfs.

As an illustrative example, consider two lumps of gas, each with mass $M/2\sim 0.015\,{\rm M}_{_\odot}$, density $n_{_{{\rm H}_2.{\rm INFLOW}}}\simeq 1.4\times 10^7\,{\rm H}_{_2}\,{\rm cm}^{-3}$ and linear size $D\sim 600\,{\rm AU}$, colliding head on at relative speed $2v_{_{\rm INFLOW}}\simeq 2\,{\rm km}\,{\rm s}^{-1}$. The gas cools rapidly to $T\sim 10\,{\rm K}$, $a\sim 0.2\,{\rm km}\,{\rm s}^{-1}$ \citep[by molecular-line cooling; e.g.][]{Whitwort2016}, and so the post-shock gas has density 
\begin{eqnarray}
n_{_{{\rm H}_2.{\rm POST-SHOCK}}}&\simeq&n_{_{{\rm H}_2.{\rm INFLOW}}}\,\left(\frac{v_{_{\rm INFLOW}}}{a}\right)^2\;\,\sim\;\, 4\times 10^8\,{\rm H}_{_2}\,{\rm cm}^{-3}\,.
\end{eqnarray}
The collision lasts $t_{_{\rm COLLISION}}\simeq D/v_{_{\rm INFLOW}}\sim 3\,{\rm kyr}$, and produces a layer with extent $D$ and half-thickness $Z\sim D \left(a/v_{_{\rm INFLOW}}\right)^2\sim 25\,{\rm AU}$. After this, the layer is marginally unstable against lateral contraction, {\it parallel} to the midplane of the layer ($Da^2/GM\sim 1$), and the instability develops on a timescale
\begin{eqnarray}
t_{_{\rm LATERAL}}&\stackrel{>}{\sim}&\left(\frac{D^3}{GM}\right)^{\!1/2}\;\,\sim\;\,17\,{\rm kyr}\,.
\end{eqnarray}
However, at the same time, because the ram-pressure of the inflow has ceased, and the layer is far from being unstable against collapse in the direction {\em perpendicular} to the midplane of the layer ($Da/(GMZ)^{1/2}\gg 1$), it expands and disperses in this direction, on a timescale 
\begin{eqnarray}
t_{_{\rm DISPERSE}}&\sim&\frac{Z}{a}\;\,\sim\;\,0.7\,{\rm kyr}\,.
\end{eqnarray}

Second, turbulent fragmentation \citep{PadoNord2002} predicts that at low masses, for every core that collapses to form a low-mass star, there are many more that simply bounce and disperse. At $0.03\,{\rm M}_{_\odot}$, there should be at least 30 cores that don't condense out as Brown Dwarfs for every one that does (Nordlund, private communication), but these have not yet been detected in sufficient numbers to check the statistics.  \citet{Andretal2012} have identified a low-mass core whose mass is probably in the Brown-Dwarf range \citep[see][]{Lomaetal2016}, and whose internal velocity dispersion is low. However, there are many other possible interpretations of the observations, in particular that it is a bouncing elongated core seen end-on.

\section{Forming Brown Dwarfs by filament fragmentation}\label{SEC:FilFrag}

In this section, we consider the formation of Brown Dwarfs in the filaments that feed matter into a forming star cluster, for example the spokes of a cluster forming in a hub-and-spoke gas flow \citep[e.g.][]{Pereetal2013,Balfetal2017}. \citet{Bateetal2002} suggest, on the basis of numerical simulations of a large turbulent star-forming cloud, that the majority of Brown Dwarfs form by disc fragmentation (see next section), but some Brown Dwarfs form in the filamentary streams that deliver matter into the centres of forming star clusters. Because these Brown Dwarfs arrive late, there are already more massive stars in the central cluster. Moreover, the Brown Dwarfs tend to arrive with large velocity relative to the cluster. Therefore either they pass straight through the centre of the cluster and out the other side, or, if they do interact dynamically with individual stars in the centre of the cluster, they may well be thrown out with even higher velocity than they came in with. As a consequence, they do not usually accrete much extra mass, and therefore they are likely to remain below the Hydrogen-Burning Limit. 

The theory of how evolving filaments fragment (as distinct from static equilibrium filaments) is very complex \citep[e.g.][]{Claretal2016}. The situation described above is complicated by the fact that the filaments in question are accumulating mass from their surroundings, and at the same time the mass in a filament is falling towards the centre of the cluster. Thus, in order to analyse the stability of the filament, one must take account of the pressure and self-gravity of the gas in the filament, the ram pressure of the matter accreting onto it, and the tidal forces exerted by the cluster, which will tend to pinch the filament perpendicular to its length, and stretch it along its length. If we assume that the pinching and stretching cancel each other out, and if we also assume that the ram pressure of the matter accreting onto the filament is delivered by gas with density $n_{_{{\rm H}_2{\rm .INFLOW}}}$ travelling at velocity $v_{_{\rm INFLOW}}$,
\begin{eqnarray}
P_{_{\rm RAM}}&\simeq&\rho_{_{\rm INFLOW}}\,v_{_{\rm INFLOW}}^2\;\,\simeq\;\,n_{_{{\rm H}_2{\rm .INFLOW}}}\,{\bar m}_{_{{\rm H}_2}}\,v_{_{\rm INFLOW}}^2\,,
\end{eqnarray}
then the filament should fragment into condensations with mass
\begin{eqnarray}
M_{_{\rm FIL.FRAG}}\!\!&\!\!\simeq\!\!&\!\!\frac{30a^4}{G^{3/2}\rho_{_{\rm INFLOW}}^{1/2}v_{_{\rm INFLOW}}}\longrightarrow0.03\,{\rm M}_{_\odot}\!\left(\frac{n_{_{{\rm H}_2{\rm .INFLOW}}}}{3\times 10^6\,{\rm cm}^{-3}}\right)^{\!-1/2}\!\left(\frac{v_{_{\rm INFLOW}}}{{\rm km}\,{\rm s}^{-1}}\right)^{\!-1}\!.\hspace{0.8cm}
\end{eqnarray}
Here, we are assuming that the turbulence in the filament is subsonic. Thus the formation of Brown Dwarfs by filament fragmentation requires the filament to be accreting rather dense gas at relatively high speed.

\section{Forming Brown Dwarfs by disc fragmentation}\label{SEC:DiscFrag}

In this section we consider the circumstances under which a protostellar disc around a primary star fragments, thereby forming low-mass secondary companions. When a core contracts to form a new star, the expectation is that the core has finite angular momentum, due to turbulence in the material that collected to form the core in the first place. Initially, only the matter with very low angular momentum flows directly into the primary star, and the rest forms an accretion disc around the primary. Here, torques due to gravity and/or magnetic fields transfer angular momentum outwards, allowing additional material from the inner disc to flow into the primary star, whilst the material in the outer disc acquires angular momentum and expands outwards. If the rate at which material with high angular momentum is delivered to the accretion disc is high, and the speed with which some of this material can then lose angular momentum and migrate inwards is low, the disc may find itself with comparable mass to the primary. Under this circumstance, and provided it is sufficiently cool, the disc fragments gravitationally to form secondary stars.

An equilibrium circularly-symmetric circumstellar disc, around a primary star with mass $M_{_\star}$, is characterised by its surface-density, $\Sigma({\cal R})$ (where ${\cal R}$ is the distance from the primary), its isothermal sound speed, $a({\cal R})$, and its angular speed, $\Omega({\cal R})$. To determine when, where and how fast the disc fragments gravitationally, we focus on a small circular patch of radius $R$, at distance ${\cal R}$ from the primary, with the proviso that $R\ll{\cal R}$, i.e. the patch is much smaller than the distance to the primary. Contraction of this patch is controlled by the balance between its self-gravity, internal pressure and spin,
\begin{eqnarray}
\ddot{R}&\simeq&-\,2\,\pi\,G\,\Sigma({\cal R})\;+\;\frac{a^2({\cal R})}{R}\;+\;\epsilon^2({\cal R})\,R\,,
\end{eqnarray}
where $\epsilon({\cal R})$ is the epicyclic frequency, given by 
\begin{eqnarray}
\epsilon^2({\cal R})&=&\frac{2\,\Omega({\cal R})}{{\cal R}}\,\frac{d}{d{\cal R}}\!\left({\cal R}^2\,\Omega({\cal R})\right)\,;
\end{eqnarray}
provided the disc is not too massive, compared with the primary , we can set 
\begin{eqnarray}
\epsilon({\cal R})&\simeq&\Omega({\cal R})\;\,\simeq\;\,\left(\frac{G\,M_{_\star}}{{\cal R}^3}\right)^{\!1/2}\,.
\end{eqnarray}

The condition for contraction is $\ddot{R}<0$, and so unstable patches have radii in the range $(R_{_{\rm MIN}},R_{_{\rm MAX}})$, where
\begin{eqnarray}\label{EQN:R_min,max}
R_{_{\rm MIN/MAX}}&\simeq&\frac{(\pi G\Sigma)\,\mp\,\left\{(\pi G\Sigma)^2\,-\,(a\Omega)^2\right\}^{1/2}}{\Omega^2}\,,
\end{eqnarray}
and we have dropped the dependence of $\Sigma$, $a$ and $\Omega$ on ${\cal R}$ for the sake of compactness. Patches with $R<R_{_{\rm MIN}}$ are unable to condense out because their pressure support is stronger than their self-gravity. Patches with $R>R_{_{\rm MAX}}$ are unable to condense out because their spin support is stronger than their self-gravity.

Eqn. (\ref{EQN:R_min,max}) only has real roots, and hence the disc can only fragment, if $(\pi G\Sigma)\,>\,(a\Omega)$, i.e.
\begin{eqnarray}\label{EQN:Toomre}
\Sigma({\cal R})&>&\Sigma_{_{\rm MIN}}({\cal R})\;\,\simeq\;\,\frac{a({\cal R})\,\Omega({\cal R})}{\pi\,G}\,.
\end{eqnarray}
This is the Toomre Condition for gravitational fragmentation of an equilibrium disc \citep{ToomreA1964}. The timescale on which an unstable patch (hereafter a proto-fragment) condenses out is given by
\begin{eqnarray}
t_{_{\rm COND}}&\simeq&\left(\frac{2\,R}{\ddot{R}}\right)^{\!1/2}\;\,\simeq\;\,\left\{\frac{\pi G\Sigma}{R}\,-\,\frac{a^2}{2R^2}\,-\,\frac{\Omega^2}{R}\right\}^{-1/2}\,.
\end{eqnarray}
The fastest condensing fragment has radius, mass and condensation time given by
\begin{eqnarray}
R_{_{\rm FASTEST}}&\simeq&\frac{a^2}{\pi G\Sigma}\,,\\
M_{_{\rm FASTEST}}&\simeq&\pi\,R_{_{\rm FASTEST}}^2\,\Sigma\;\simeq\;\frac{a^4}{\pi\,G^2\,\Sigma}\,,\\
t_{_{\rm FASTEST}}&\simeq&\left\{\frac{(\pi G\Sigma)^2}{2a^2}\,-\,\frac{\Omega^2}{2}\right\}^{-1/2}\;\,\simeq\;\,\frac{t_{_{\rm ORBIT}}}{2^{1/2}\pi\left\{(\Sigma/\Sigma_{_{\rm MIN}})^2\,-\,1\right\}^{1/2}}\,,
\end{eqnarray}
where
\begin{eqnarray}
t_{_{\rm ORBIT}}&=&\frac{2\pi}{\Omega}
\end{eqnarray}
is the orbital period at radius ${\cal R}$. It follows that a proto-fragment can condense out in one orbital period if 
\begin{eqnarray}
\frac{\Sigma}{\Sigma_{_{\rm MIN}}}&\stackrel{>}{\sim}&1\,+\,\frac{1}{(2\pi)^2}\;\,\simeq\;\,1.025\,;
\end{eqnarray}
in other words, the disc need only be marginally unstable for proto-fragments to start condensing out on a dynamical timescale.

As with a prestellar core, contraction only continues if the proto-fragment can radiate away, on a dynamical timescale, the compressional work being done on the gas. Otherwise the gas heats up, the proto-fragment bounces and expands, and it is then sheared apart. 

The rate of radiative cooling of the proto-fragment, from the two sides of the disc, is
\begin{eqnarray}
{\cal C}_{_{\rm RAD.FASTEST}}&\simeq&\frac{2\,\pi\,R_{_{\rm FASTEST}}^2\,\sigma_{_{\rm SB}}\,T^4}{\left({\bar \tau}_{_{\rm ROSSELAND}}\,+\,{\bar \tau}_{_{\rm PLANCK}}^{-1}\right)}\;\,\simeq\;\,\frac{2^2\,(\pi\,{\bar m})^4\,a^8}{15\,\kappa_{_a}\,(c\,G)^2\,(h\Sigma)^3}\,,
\end{eqnarray}
where this time we have assumed that the proto-fragment is optically thick to its own cooling radiation, and so we have included the term ${\bar\tau}_{_{\rm ROSSELAND}}\simeq\Sigma\kappa_{_a}a^4$. The rate of compressional heating of the contracting proto-fragment is given by
\begin{eqnarray}
{\cal H}_{_{\rm COMP.FASTEST}}\!&\!\simeq\!&\!\left(P\frac{dV}{dt}\right)_{_{\rm FASTEST}}\simeq\frac{P_{_{\rm FASTEST}}V_{_{\rm FASTEST}}}{t_{_{\rm ORBIT}}}\simeq\frac{M_{_{\rm FASTEST}}a^2}{t_{_{\rm ORBIT}}}\simeq\frac{a^6\Omega}{2\pi^2G^2\Sigma}\,.\hspace{0.7cm}
\end{eqnarray}
and the requirement that ${\cal C}_{_{\rm RAD.FASTEST}}\!\stackrel{>}{\sim}\!{\cal H}_{_{\rm COMP.FASTEST}}$ reduces to an upper limit on $\!\Sigma\!,$
\begin{eqnarray}\label{EQN:Sigma_max}
\Sigma&\stackrel{<}{\sim}&\Sigma_{_{\rm MAX}}\;\,\simeq\;\,\left(\frac{8}{15\,h^3\,\kappa_{_a}\,\Omega}\right)^{\!1/2}\,\frac{\pi^3\,{\bar m}^2\,a}{c}\,.
\end{eqnarray}

If the disc is to fragment, this upper limit ($\Sigma_{_{\rm MAX}}$; Eqn, \ref{EQN:Sigma_max}) must be greater than the lower limit for gravitational instability ($\Sigma_{_{\rm MIN}}$; Eqn. \ref{EQN:Toomre}), which yields
\begin{eqnarray}
\Omega&\stackrel{<}{\sim}&\left(\frac{8\,\pi^8\,G^2\,{\bar m}^4}{15\,c^2\,h^3\,\kappa_{_a}}\right)^{\!1/3}\,.
\end{eqnarray}
Finally, an upper limit on $\Omega$ corresponds to a lower limit on the radius at which disc fragmentation can occur,
\begin{eqnarray}\label{EQN:R_min}
{\cal R}&\stackrel{>}{\sim}&{\cal R}_{_{\rm MIN}}\;\,\simeq\;\,\left(\frac{15\,c^4\,h^6\,\kappa_{_a}^2\,M_{_\star}^3}{2^6\,\pi^{16}\,G\,{\bar m}^8}\right)^{1/9}\;\,\simeq\;\,70\,{\rm AU}\left(\frac{M_{_\star}}{{\rm M}_{_\odot}}\right)^{\!1/3}\,.
\end{eqnarray}
Since the fragment is assumed to be optically thick, we must have
\begin{eqnarray}
{\bar\tau}_{_{\rm ROSSELAND}}&\simeq&\Sigma\kappa_{_a}a^4\;\,\stackrel{>}{\sim}\;\,1\,,\\
M_{_{\rm FASTEST}}&\simeq&\frac{a^4}{\pi\,G^2\,\Sigma}\;\,\stackrel{<}{\sim}\;\,\frac{\kappa_{_a}\,a^8}{\pi\,G^2}\;\,\simeq\;\,0.1\,{\rm M}_{_\odot}\left(\frac{T}{30\,{\rm K}}\right)^4\,.
\end{eqnarray}
We conclude that disc fragmentation is only likely to occur in the outer parts of protostellar discs (Eqn. \ref{EQN:R_min}), and since the temperatures here are $T\stackrel{<}{\sim}30\,{\rm K}$, the initial masses are predominantly in the Brown-Dwarf range. Disc fragmentation will often lead to the formation of more than one low-mass companion, and interactions between companions will tend to eject some, whilst others are scattered into the inner disc. The ones that are ejected, if they are ejected quite quickly, have not had much time to accrete more matter, so they will tend to retain their low initial masses and populate the field with free-floating Brown Dwarfs. Conversely, the ones that stay in the disc will accrete additional matter from the disc, and migrate inwards towards the primary star, becoming secondary companions to the primary star. This explains the Brown Dwarf Desert, the observed paucity of Brown-Dwarf companions in close orbits around Sun-like stars; by the time a secondary companion has migrated in to a close orbit it has acquired additional mass from the residual disc and become a hydrogen-burning star.  A few Brown Dwarfs may remain on wide orbits, but the majority end up in the field, either as a consequence of being dynamically ejected, or as a consequence of being liberated by the tidal stresses to which they are repeatedly subjected (due to passing more massive stars, star clusters and molecular clouds).

\begin{figure}
\includegraphics[scale=.65]{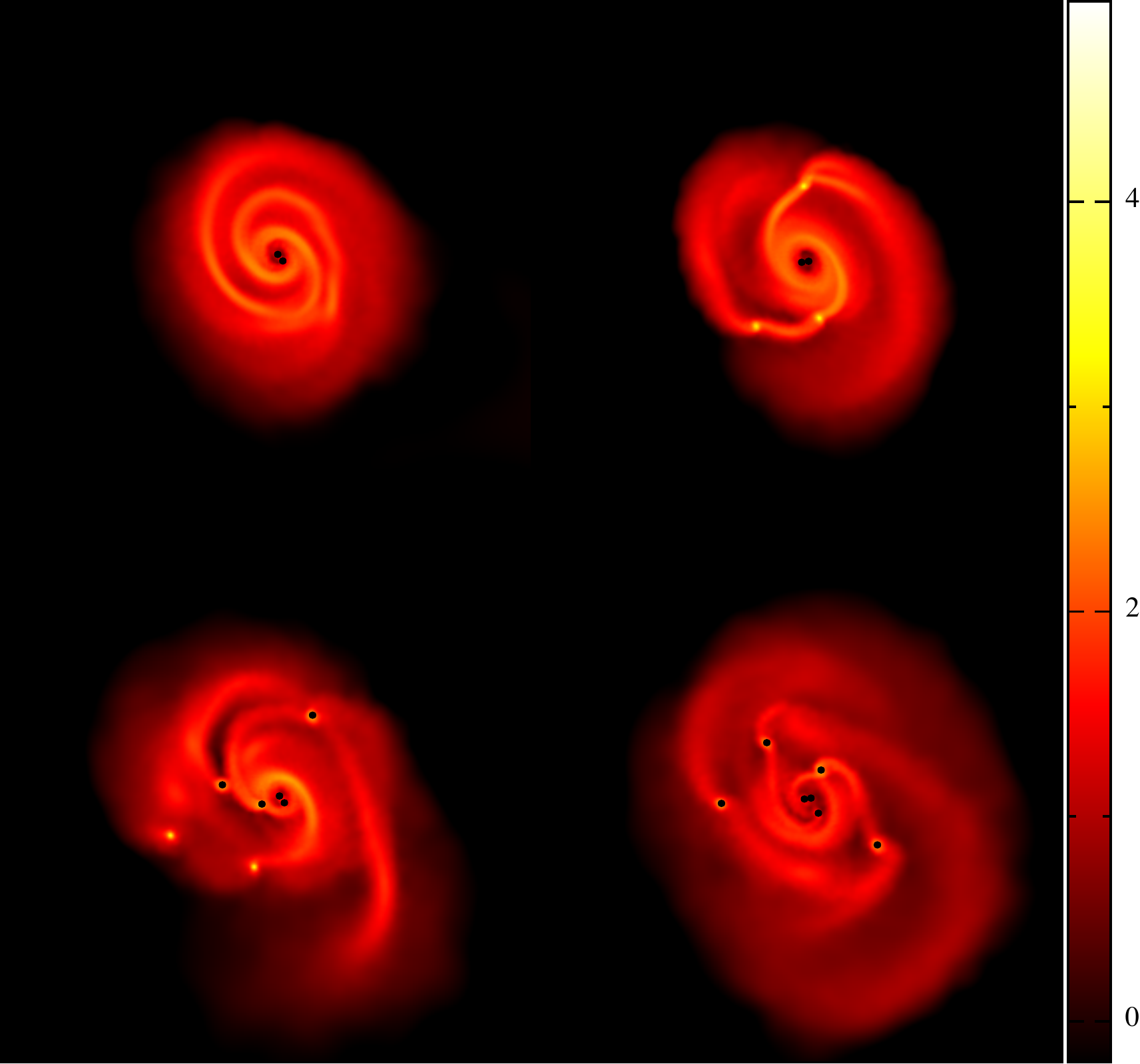}
\caption{This montage shows four frames from a simulation of a collapsing prestellar core, with initial conditions matching the cores observed in Ophiuchus. The frames are at times $t=59,\,60,\,61\;{\rm and}\;62\;{\rm kyrs}$, and show the formation of five Brown Dwarfs in a circumbinary disc around a close binary system comprising two M dwarfs. The disc is approximately $300\,{\rm AU}$ across, and the colour bar gives the column-density in ${\rm g}\,{\rm cm}^{-2}$. (Courtesy of Dr. Oliver Lomax.)}
\label{FIG:discfrag}
\end{figure}

There have been many simulations of the formation of Brown Dwarfs by disc fragmentation, following the collapse of a prestellar core, in particular those by \citet{Stametal2007,Stametal2009,Stametal2011} and \citet{Lomaetal2014,Lomaetal2015,Lomaetal2016}, which use initial conditions matching, in a statistical sense, the cores observed in Ophiuchus \citep{Mottetal1998,Andretal2007}. These simulations suggest that Brown Dwarfs can form by disc fragmentation, in the numbers observed and with the properties observed, but only if (i) a significant fraction of the turbulent energy in the initial core is in solenoidal modes (at least $\sim 30\%$, which is likely, because the natural/thermal fraction is $67\%$); and (ii) radiative feedback from the primary star (and any other stars formed subsequently) is episodic, with a duty cycle measured in kiloyears -- as has been inferred observationally by \citet{Schoetal2013} and predicted on the basis of a phenomenological model by \citet{ZhuZetal2010}. Fig. \ref{FIG:discfrag} shows five Brown Dwarfs formed in a circumbinary disc in one of these simulations.

\section{Forming Brown Dwarfs by dynamical ejection}\label{SEC:DynEject}

In this section, we consider the formation of Brown Dwarfs when stellar embryos are prematurely ejected from their birth clusters, and consequently accretion is terminated before they have grown to sufficient mass to form hydrogen-burning stars. \citet{ReipClar2001} have pointed out that, in a cluster of newly-formed stars, dynamic relaxation tends to lead to the ejection of the lower-mass stars, and they point out that this may be a critical process in the formation of Brown Dwarfs and their delivery into the field. They distinguish two basic scenarios.

In the first scenario, a small cluster or subcluster, of the sort that forms in the strongly shock-compressed layer that is formed when clouds collide at {\em high} speed \citep{Balfetal2015}, spawns a small number of stars in a tight configuration with linear size $\stackrel{<}{\sim} 100\,{\rm AU}$. Unless interactions between the stars are moderated by the viscous drag of attendant discs \citep{McDoClar1995}, dynamical interactions between the stars lead to the ejection of the least massive ones \citep{McDoClar1993}, and for reasonable parameters, \citet{ReipClar2001}  show that this is likely to happen before these low-mass stars are massive enough to ignite hydrogen burning. In this scenario, the ejection speeds are expected to be of order $\sim 3\,{\rm km}\,{\rm s}^{-1}$. Therefore one might expect to find a diaspora of Brown Dwarfs in the immediate vicinity of  Class 0 protostars and subsequently in the outskirts of young clusters like the Pleiades. 

In the second scenario, a more monolithic cluster, of the sort that forms in the weakly shock-compressed layer that is formed when clouds collide at {\em low} speed \citep{Balfetal2015}, spawns a large number of stars in a more extended configuration, with linear size $\stackrel{>}{\sim} 10000\,{\rm AU}$. Under this circumstance, the final masses of the stars are largely determined by competitive accretion \citep{Bonnetal1997}. Stars that end up trapped at the bottom of the cluster's gravitational potential well, enjoy a copious supply of material to accrete, and many grow to large mass. Conversely, stars that end up on extended orbits in the cluster's gravitational potential tend to be, and to remain, low-mass; they spend most of their time in the outer reaches of the cluster, where there is little material for them to accrete, and if they occasionally pass through the dense central zone of the cluster where there is dense gas to accrete, they are moving so fast that they accrete little of it. In fact, even if a low-mass star ends up moving slowly in the central zone, it may well undergo a three-body interaction and get kicked out of the cluster all together; the energy gained by the low-mass star is probably at the expense of a binary system involving more massive components which then end up more tightly bound than before. These ejected low-mass stars are likely to include Brown Dwarfs, and most will leave with low velocities, $\sim 0.3\,{\rm km}\,{\rm s}^{-1}$. A further consideration that will facilitate the formation of Brown Dwarfs in this second scenario arises if there is a spread in the formation times of stars; those that form late on find themselves in a situation where there is little gas left to accrete, so they are more likely to end up as Brown Dwarfs.

\section{Forming Brown Dwarfs by photo-erosion}\label{SEC:PhotoEro}

In this section we consider the possibility that Brown Dwarfs form when a core is overrun by an H{\sc ii} region \citep{Hestetal1996}. The scenario envisaged here is that a cluster of OB stars forms in a massive molecular cloud, and rapidly ionises the surrounding gas to form an H{\sc ii} region. The act of ionisation increases the temperature of the gas by a factor of $\sim 300$ (from $\sim 30\,{\rm K}$ to $\sim 10^4\,{\rm K}$) and converts each hydrogen molecule into four particles (two protons and two electrons); the combination of these two effects increases the pressure by a factor of order $\sim 10^3$. If the ionisation front encounters a massive core that is significantly denser than its surroundings but not as yet gravitationally unstable (so it is not already contracting to form a star), the advance of the ionisation front is slowed, and the H{\sc ii} region wraps round the core. The core is now exposed to direct ionising radiation from the OB stars, and to the diffuse ionising radiation from recombination in the H{\sc ii} region. Consequently the core is surrounded by an ionisation front which steadily eats into it, boiling off ionised gas as it goes. At the same time, the advancing ionisation front is preceded by a shock front, which compresses the inner regions of the core, and this may be sufficient to cause them to condense and form a star. The final mass of this star is determined by a competition between the rate at which the ionisation front boils off the outer layers of the core and the rate at which the inner regions contract to stellar density.

This mechanism has been analysed by \citet{WhitZinn2004}, who argue that there are three consecutive phases involved. In the first phase, the compression wave ahead of the ionisation front converges on the centre of the core. If the density in the ionised gas surrounding the core is $\rho_{_{\rm O}}$, this first phase takes a time
\begin{eqnarray}
t_{_1}&\simeq&\frac{1}{(2\pi G\rho_{_{\rm O}})^{1/2}}\,.
\end{eqnarray}
At the end of this first phase, the compression wave reaches the centre and forms a protostar. Thereafter the protostar accretes at a rate
\begin{eqnarray}\label{EQN:Mdot}
{\dot M}_{_\star}&\simeq&\frac{a_{_{\rm I}}^3}{G}\,,
\end{eqnarray}
where $a_{_{\rm I}}\simeq 0.3\,{\rm km}\,{\rm s}^{-1}$ is the isothermal sound speed in the neutral gas; this is typically a little higher than in low-mass star formation regions, due to the proximity of the OB stars.

In the second phase, an expansion wave propagate outwards from the protostar at speed $a_{_{\rm I}}$, so its radius is
\begin{eqnarray}
r_{_{\rm EW}}&\simeq&a_{_{\rm I}}(t-t_{_1})\,.
\end{eqnarray}
and interior to the expansion wave the material flows inwards onto the protostar, approximately in free fall, i.e.
\begin{eqnarray}
\rho(r)&\simeq&\frac{3\,a_{_{\rm I}}^2}{8\,\pi\,G\,r_{_{\rm EW}}^{1/2}\,r^{3/2}}\,,\\
v(r)&\simeq&\frac{2\,a_{_{\rm I}}\,r_{_{\rm EW}}^{1/2}}{3\,r^{1/2}}\,,
\end{eqnarray}
feeding the accretion rate given in Eqn. (\ref{EQN:Mdot}). This  second phase is relatively short, compared with the first one, and ends when the outward propagating expansion wave meets the inward propagating ionisation front at time $t_{_2}$.

In the third phase, which is also quite short, the ionisation front continues to propagate inwards, but now it is overtaking material inside the expansion wave which is therefore falling inwards towards the central protostar. Ionised gas flows off the back of the ionisation front at speed $v\simeq (5/3)^{1/2}a_{_{\rm II}}$, relative to the ionisation front, where $a_{_{\rm II}}\simeq 10\,{\rm km}\,{\rm s}^{-1}$ is the isothermal sound speed in the ionised gas at $T\simeq 10^4\,{\rm K}$. The third phase ends at $t=t_{_3}$, when the specific kinetic energy of the matter flowing off the back of the ionisation front, relative to the protostar,
\begin{eqnarray}
\frac{({\dot r}_{_{\rm IF}}+(5/3)^{1/2}a_{_{\rm II}})^2}{2}\,,
\end{eqnarray}
becomes less than its gravitational binding energy
\begin{eqnarray}
\frac{G(M_{_\star}+M_{_{\rm IF}})}{r_{_{\rm IF}}}\,;
\end{eqnarray}
here $r_{_{\rm IF}}$ is the radius of the ionisation front, ${\dot r}_{_{\rm IF}}\equiv dr_{_{\rm IF}}/dt$, $M_{_\star}$ is the mass of the protostar, and $M_{_{\rm IF}}$ is the mass of infalling matter interior to the ionisation front. The assumption is that, for $t<t_{_3}$, all the matter flowing off the ionisation front escapes, whereas all the matter interior to the ionisation front at $t=t_{_3}$ is accreted onto the protostar; even with the velocity boost delivered by the ionisation front, this matter is unable to escape. Hence the final mass of the protostar is given by 
\begin{eqnarray}
M_{_{\rm FINAL}}&\simeq&M_{_\star}(t_{_3})\,+\,M_{_{\rm IF}}(t_{_3})\,.
\end{eqnarray}
For a wide range of parameters, $M_{_\star}(t_{_3})$ and $M_{_{\rm IF}}(t_{_3})$ turn out to be comparable, and the final mass is given by
\begin{eqnarray}
M_{_{\rm FINAL}}&\simeq&0.03\,{\rm M}_{_\odot}\,\left(\frac{a_{_{\rm I}}}{0.3\,{\rm km}\,{\rm s}^{-1}}\right)^{\!6}\,\left(\frac{\dot{\cal N}_{_{\rm LyC}}}{10^{50}\,{\rm s}^{-1}}\right)^{\!-1/3}\,\left(\frac{\rho_{_{\rm O}}}{10^{-22}\,{\rm g}\,{\rm cm}^{-3}}\right)^{\!-1/3}\,,\hspace{0.8cm}
\end{eqnarray}
so this mechanism probably does operate to form Brown Dwarfs in regions of high-mass star formation; it is probably the cause of the globulettes observed by \citet{Gahmetal2007}. However, it is unlikely to be a major formation mechanism for Brown Dwarfs. First, it is very inefficient. It requires a very massive core to form a single Brown Dwarf, 
\begin{eqnarray}
\frac{M_{_{\rm FINAL}}}{M_{_{\rm CORE}}}&\simeq&10^{-4}\,\left(\frac{M_{_{\rm CORE}}}{{\rm M}_{_\odot}}\right)\,\left(\frac{R_{_{\rm HII}}}{\rm pc}\right)^{\!-1}\,,
\end{eqnarray}
where $M_{_{\rm CORE}}$ is the initial mass of the core and $R_{_{\rm HII}}$ is the radius of the H{\sc ii} region. Second, it can only operate where there are OB stars to deliver ionising radiation, but there are many Brown Dwarfs observed in low-mass star formation regions where there are no OB stars.

\section{Discussion}\label{SEC:Caveats}

In order to bring out the basic physics of Brown Dwarf formation, we have where possible relied on analytic arguments, rather than computer simulations, provided these can capture the basic physics and the trends that result from it. This has involved invoking a one-zone model of a core, even though we know that, as a core becomes denser, the contrast between conditions in the centre and at the edge tends to increase, and therefore accurate results can only be obtained with computer simulation. Moreover, since the process of star formation entails the interplay of non-linear processes, it is chaotic, and collective predictions can only be made by performing detailed numerical simulations and following many different realisations. Simulations of the processes that might be involved in forming Brown Dwarfs are still in their infancy, and for some of the mechanisms proposed in the preceding sections have not been attempted at all. There is much work to be done.

We have also avoided discussing at length, or invoking, constraints based on observation, except where we feel that the constraint is solid (i.e. the existence of Brown Dwarfs, the presence of Brown Dwarfs in star formation regions where there have never been any ionising stars, and the Brown Dwarf Dessert). There are other constraints which may eventually help in identifying the main mechanism or mechanisms that form Brown Dwarfs. However, the statistics are poor, and corrupted by selection effects. Furthermore, as with hydrogen-burning stars, one has to distinguish the statistics of young populations (which probably still bear signatures of their formation mechanism but are the hardest to acquire, due to being embedded in their birth clouds) from the statistics of field populations (which are somewhat easier to acquire, but less likely to relate closely to the birth environment). This distinction is particularly important for Brown Dwarfs, because, by virtue of their low masses, they are particularly susceptible to the tidal impulses that destroy multiple systems and remove discs, and hence delete signatures of the birth environment.

The Initial Mass Function appears to be consistent with a log-normal form, extending to Brown-Dwarf masses, but this result is very noisy, because it is hard to obtain accurate masses of Brown Dwarfs --- and it becomes ever noisier as one proceeds to Brown Dwarfs with the lowest masses, where it eventually becomes confused with the mass function of planets.

The binary statistics of Brown Dwarfs are again consistent with being an extension of those for low-mass hydrogen-burning stars. In particular, the fraction of Brown Dwarfs that are primaries in binary systems decreases with mass, as it appears to do over the entire range of stellar masses. Some additional features of binary systems that include Brown Dwarfs have been noted. First, systems with Brown Dwarf primaries seem usually to be close systems, with most separations being $\stackrel{<}{\sim} 20\,{\rm AU}$. Second, brown dwarfs that are companions to more massive stars (in particular, spectral types G, K and M) tend to be on quite wide orbits, $\stackrel{>}{\sim} 70\,{\rm AU}$. Third,  there is tentative evidence that these Brown Dwarfs (the ones in orbit around a more massive star) have a higher likelihood of being in a close binary system with another Brown Dwarf than do Brown Dwarfs in the field. In other words, these are hierarchical  triple systems, in which a tight BD+BD binary (separation of order 5 to $10\,{\rm AU}$) is on a wide orbit  about a Sun-like star (separation of order 100 to $200\,{\rm AU}$).

Many young Brown Dwarfs have accretion discs and show the emission lines characteristic of ongoing accretion that are seen in hydrogen-burning stars. The estimated accretion rates also appear to follow the same approximate scaling relation, $\dot{M}\propto M^2$, but with quite large scatter.  

The binary statistics, and the evidence for accretion discs, have sometimes been invoked to dismiss the possibility that Brown Dwarfs in the field can have been ejected from their birth site, but this is fallacious. Ejection velocities are in general quite low, and simulations indicate that ejected Brown Dwarfs can involve BD/BD binaries that stay intact, and accretion discs that are truncated but not destroyed. Indeed, the statistics of BD/BD binaries suggest that they may well be formed in discs, but these statistics need to be consolidated before this conclusion can be drawn with any confidence.

\section{Conclusions}\label{SEC:Conc}

This review is predicated on the presumption that, as regards their formation,  Brown Dwarfs should not be distinguished from hydrogen-burning stars. As one proceeds across the Hydrogen-Burning Limit to lower masses, there might be a steady shift away from a mix dominated by one formation scenario (say formation by dynamical fragmentation) towards a mix dominated by another formation scenario (say disc fragmentation), but there is no abrupt change in the mix. We have therefore treated Brown Dwarfs as very low-mass stars, which form on a short, dynamical timescale, by gravitational instability and with an initially homogeneous elemental composition. This distinguishes them from planets, which we envisage as objects that form by core accretion, on a much longer timescale and with a fractionated elemental composition. There are also hybrid formation scenarios, but as yet these are not sufficiently well developed to merit inclusion in this short review. 

With this distinction between stars and planets in mind, we have listed and discussed the minimum mass for a Brown Dwarf (effectively the Opacity Limit for star formation); the maximum mass for a Brown Dwarf (effectively the Hydrogen-Burning Limit); and five mechanisms that might be involved in the formation of Brown Dwarfs. These are turbulent fragmentation, filament fragmentation, disc fragmentation, dynamical ejection and photo-erosion. These mechanisms are not all mutually exclusive. In particular, filament fragmentation and disc fragmentation are probably sometimes followed by dynamical ejection. Photo erosion probably occurs occasionally, but it is unlikely to be a major source of Brown Dwarfs. The critical distinction would then seem to be between, on the one hand, turbulent fragmentation, and on the other hand filament and/or disc fragmentation, followed sometimes by dynamical ejection. In this context, the recent discovery by \citet{Tobietal2016} of what appears to be a Brown Dwarf, embedded in a protostellar disc, is exciting, but without many more examples this cannot be used to discriminate against turbulent fragmentation.

\begin{acknowledgement}
APW gratefully acknowledges the support of the UK's Science and Technology Facilities Council, through Consolidated Grant ST/K00926/1.
\end{acknowledgement}

\bibliographystyle{spbasicHBexo}  
\bibliography{Whitworth} 

\end{document}